\documentclass{appolb}

\def\slashchar#1{\setbox0=\hbox{$#1$}
   \dimen0=\wd0 \setbox1=\hbox{/} \dimen1=\wd1
   \ifdim\dimen0>\dimen1 \rlap{\hbox to \dimen0{\hfil/\hfil}} #1
   \else  \rlap{\hbox to \dimen1{\hfil$#1$\hfil}} / \fi}

\usepackage{cite}
\usepackage{epsfig}

\begin{document}

\title{\bf Renormalization and universality of NN interactions in
  Chiral Quark and Soliton Models ~\footnote{Talk presented by ERA at
    the Mini-Workshop Bled 2009: {\it Problems in Multi-quark States}.
    Bled (Slovenia), June 29 - July 6, 2009.}}

\author{E. Ruiz Arriola and A. Calle Cord\'on
\address{Departamento de F\'{\i}sica At\'omica, Molecular y Nuclear, \\ Universidad de Granada, E-18071 Granada, Spain}}

\maketitle

\vspace{-4mm}

\abstract{We use renormalization as a tool to extract universal
  features of the NN interaction in quark and soliton nucleon models,
  having the same long distance behaviour but different short distance
  components.  While fine tuning conditions in the models make
  difficult to fit NN data, the introduction of suitable
  renormalization conditions supresses the short distance
  sensitivity. Departures from universality are equivalent to
  extracting information on the model nucleon structure.}

\vspace{5mm}

\section{Introduction}

The meson exchange picture has played a key role in the development of
Nuclear Physics~\cite{Machleidt:1987hj,Nagels:1977ze}. However, the
traditional difficulty has been a practical need to rely on short
distance information which is hardly accessible directly but becomes
relevant when nucleons are placed off-shell. From a theoretical point
of view this is unsatisfactory since one must face uncertainties not
necessarily linked to our deficient knowledge at long distances and
which are difficult to quantify. On the other hand, the purely field
theoretical derivation yields potentials which present short distance
singularities, thereby generating ambiguities even in the case of the
widely used One Boson Exchange (OBE) potential. Consider, for
instance, the venerable One Pion Exchange (OPE) $NN \to NN$ potential
which for $r\neq 0$ reads
\begin{eqnarray}
V^{1\pi}_{NN,NN} (r) 
&=& \tau_1 \cdot \tau_2 \sigma_1 \cdot \sigma_2 W_S^{1\pi} (r) +
\tau_1 \cdot \tau_2 S_{12} W_T^{1\pi}(r) \, , 
\label{eq:pot-OPE}
\end{eqnarray} 
where the tensor operator $ S_{12} = 3 \sigma_1 \cdot \hat x \sigma_2
\cdot \hat x - \sigma_1 \cdot \sigma_2 $ has been introduced and
\begin{eqnarray}
W_S^{1\pi} (r) = \frac{m_\pi}{3} \frac{f_{\pi NN}^2}{4\pi} Y_0 (m_\pi r) 
\, \quad , \qquad \, 
W_T^{1\pi} (r) =
\frac{m_\pi}{3} \frac{f_{\pi NN}^2}{4 \pi} Y_2(m_\pi r) \, .  
\end{eqnarray} 
Here $Y_0(x)=e^{-x}/x $ and $Y_2(x)=e^{-x}/x (1+3/x+3/x^2)$ and
$f_{\pi NN}= m_\pi g_{\pi NN} /(2 M_N) $; $f_{\pi
  NN}^2/(4\pi)=0.07388$ for $g_{\pi NN}=13.08$. As we see, the OPE
potential presents a $1/r^3$ singularity, but it can be handled
unambiguously mathematically and with successful deuteron
phenomenology~\cite{PavonValderrama:2005gu}. Nonetheless, the standard
way out to {\it avoid} the singularities in this and the more general
OBE case is to implement vertex functions for the meson-baryon-baryon
coupling ($mAB$) in the OBE potentials. This correspondins to a
folding in coordinate space which in momentum space becomes the
multiplicative replacement
\begin{eqnarray}
V_{m AB}(q) \to V_{m AB}(q) \left[\Gamma_{m AB}(q^2) \right]^2 \, . 
\end{eqnarray} 
where $q^2= q_0^2 - \vec q^2$ is the 4-momentum. Standard choices are
to take form factors of the mono-pole~\cite{Machleidt:1987hj} and
exponential~\cite{Nagels:1977ze} parameterizations
\begin{eqnarray}
\Gamma_{mNN}^{\rm mon} (q^2) = \frac{\Lambda^2-m^2}{\Lambda^2-q^2} \, , \quad \,  \Gamma_{mNN}^{\rm exp}(q^2) =
\exp \left[\frac{q^2-m^2}{\Lambda^2}\right] \, , \label{eq:formfac}
\end{eqnarray} 
fulfilling the normalization condition $\Gamma_{mNN}(m^2)=1$. Due to
an extreme fine-tuning of the interaction, mainly in the $^1S_0$
channel, OBE potential models have traditionally needed a too large
$g_{\omega NN} $ to overcome the mid range attraction implying one of
the largest ($\sim 40 \%$) $SU(3)$ violations known to date. In our
recent
works~\cite{RuizArriola:2007wm,CalleCordon:2008eu,CalleCordon:2008cz,CalleCordon:2009ps,RuizArriola:2009bg,Cordon:2009pj}
we discuss how this problem may be circumvented with the help of
renormalization ideas which upon imposing short distance insensitivity
sidestep the fine tuning problem and allow natural $SU(3)$ values to
be adopted in such a way that form factors and heavy mesons play a
more marginal role. Contrarily to what one might naively think,
renormalization {\it reduces} the short distance dependence provided,
of course, removing the cut-off and the imposed renormalization
conditions are mutually compatible operations.

Of course, the extended character of the nucleon as a composite and
bound state of three quarks has motivated the use of microscopic
models of the nucleon to provide an understanding of the short range
interaction besides describing hadronic spectroscopy; quark or soliton
models endow the nucleon with its finite size and incorporate basic
requirements from the Pauli principle at the quark level or as
dictated by the equivalent
topology~\cite{Oka:1984yw,AlvarezEstrada:1986wq,
  Walhout:1992ek,Valcarce:2005em}. While much effort has been invested
into determining the short range interactions, there is a plethora of
models and related approximations; it is not obvious {\it what}
features of the model are being actually tested. In fact, $NN$ studies
set the most stringent nucleon size oscillator constant value
$b_N=0.518 {\rm fm}$~\cite{Valcarce:2005em} from S-waves and deuteron
properties which otherwise could be in a wider range $b_N=0.4-0.6 {\rm
  fm}$. This shows that quark models also suffer from a fine tuning
problem. In this contribution we wish to focus on the common and
universal patterns of the various approaches and to show how these
fine tunings can be reduced to a set of renormalization conditions.

\section{The relevant scales}

From a fundamental point of view the NN interaction should be obtained
as a natural solution of the 6-q system.  However, in order to
describe the NN interaction it is far more convenient to study two 3-q
clusters with nucleon quantum numbers, a procedure also applied in
recent lattice QCD investigations of the nuclear
force~\cite{Ishii:2006ec,Aoki:2009ji}.  NN scattering in the elastic
region corresponds to resolve distances about the minimal de Broglie
wavelength associated to the first inelastic pion production
threshold, $NN \to NN \pi$, and corresponds to take $ 2 E_{\rm CM} = 2
M_N + m_\pi$ yielding $p_{\rm CM}= \sqrt{m_\pi M_N} = 360 {\rm MeV}$
which means $\lambda_{\rm min} \sim 1/\sqrt{m_\pi M_N} =0.5 {\rm fm}$.
This scale is smaller than $1\pi$ and $2\pi$ exchange (TPE) with
Compton wavelengths $1.4$ and $0.7 {\rm fm}$ respectively.  Other
length scales in the problem are comparable and even shorter namely 1)
Nucleon size, 2) Correlated meson exchanges and 3) Quark exchange
effects. All these effects are of similar range and, to some extent,
redundant. In a quark model the constituent quark mass is related to
the Nucleon and vector meson masses through $M_q = M_N/N_c =M_V/2 $
which for $N_c=3$ colours gives the estimate $M_q=310-375 {\rm
  MeV}$. Exchange effects due to e.g. One-Gluon-Exchange are $\sim
e^{-2M_q r}$ since they correspond to the probability of finding a
quark in the opposite baryon. This follows from complete Vector Meson
Dominance (for a review see e.g. \cite{O'Connell:1995wf}), which for
the isoscalar baryon density, $\rho_B(r)$, and assuming independent
particle motion yields
\begin{eqnarray}
\int d^3 x e^{i q \cdot x} \langle N | \rho_B (x) | N \rangle 
=  4 \pi \int_0^\infty \, dr \, r^2 |\phi (r)|^2 j_0 (qr) 
\sim 
\frac{M_V^2}{M_V^2+q^2}
\end{eqnarray}
suggesting a spectroscopic factor $\phi(r) \sim  e^{-M_V
  r/2} M_V /\sqrt{4\pi r}$ at large distances. As we have said and we will
discuss below these effects are somewhat marginal but if they ought to
become visible they should reflect the correct asymptotic
behaviour. In the constituent quark model the CM motion can be easily
extracted assuming harmonic oscillator wave functions, $\phi(r) \sim
e^{-b^2
  r^2/2}$~\cite{Oka:1984yw,AlvarezEstrada:1986wq,Valcarce:2005em}
which yield Gaussian form factors falling off {\it much faster} than
the experimental ones. Skyrme models without vector mesons yield
instead topological Baryon densities $\rho_B(r) \sim e^{-3 m_\pi
  r}/r^7$\cite{Walhout:1992ek} corresponding to the outer pion cloud
contributions which are longest range but pressumably yield {\it only}
a fraction of the radius. In any case quark-exchange looks very much
like direct vector meson exchange potential which is $\sim e^{-M_V
  r}$.

\section{Chiral quark soliton model}

Most high precision NN potentials providing $\chi^2/{\rm DOF} < 1 $
need to incorporate universally the One-Pion-Exchange (OPE) potential
(including charge symmetry breaking effects) while the shorter range
is described by many and not so similarly looking
interactions~\cite{Stoks:1994wp}. This is probably a confirmation that
chiral symmetry is spontaneously broken at longer distances than
confinement, since hadronization has already taken place. It also
suggests that in a quark model aiming at describing NN interactions
the pion must be effectively included. Chiral quark models accomplish
this explicitly under the assumption that confinement is not crucial
for the binding of $\pi$, $N$ and $\Delta$. Pure quark models
including confinement or not have to face in addition the problem of
recovering the pion from quark-gluon dynamics. In between, hybrid
models have become practical and
popular~\cite{Oka:1984yw,AlvarezEstrada:1986wq,Valcarce:2005em}.  As
mentioned, all these scales around the confinement scale are mixed
up. Because these effects are least understood and trigger side
effects such as spurious colour Van der Waals forces arising from
Hidden color singlet states $[{\bf 88}]_A$
states~\cite{Gavela:1979zu,Greenberg:1981xn} in the (presumably
doubtful) adiabatic approximation, we will cavalierly ignore the
difficulties by remaining in a regime where confinement is not
expected to play a role and stay with standard chiral quark models.

While both the constituent chiral quark model and the Skyrme soliton
model look very disparate the Chiral Quark Soliton Model embeds both
models in the small and the large soliton limit
respectively~\footnote{Within the large $N_c$ framework the difference
  corresponds to a saddle point approximation around a trivial or
  non-trivial background. The question {\it which } regime is the
  appropriate one is a dynamical
  issue~\cite{Christov:1995vm,Weigel:2008zz}. Likewise, when the
  soliton is large, quarks are deeply bound and the topological
  soliton picture of Skyrme sets in, giving the appearance of a
  confined state (where colour Van der Waals forces cannot take
  place). The soliton of the Spectral Quark model does not allow this
  interpretation as baryon charge is never
  topological~\cite{Arriola:2006ds}.}.  We analyze the intuitive
non-relativistic chiral quark model (NRCQM) explicitly and comment on
the soliton case where similar patterns emerge. The comparison
stresses common aspects of the quark soliton model pictures which
could be true features of QCD.  While the long distance universality
between both NRCQM and Skyrme soliton model NN calculations may appear
somewhat surprising this is actually so because in a large $N_c$
framework both models are just different realizations of the
contracted spin-flavour symmetry~\cite{Kaplan:1996rk}.

\section{The non-relativistic chiral quark model}

To fix ideas it is instructive to consider the chiral-quark model
which corresponds to the Gell-Mann--Levy sigma model Lagrangean at the
quark level~\cite{Birse:1983gm} (the non-linear version 
suggested in Ref.~\cite{Manohar:1983md} will be discussed below),
\begin{eqnarray} 
{\cal L}= \bar q \left( i \slashchar{\partial} - g_{\pi q q}
(\sigma + i \gamma_5 \tau \cdot \pi ) \right) q + \frac12 \left[ 
(\partial^\mu \sigma)^2 + (\partial^\mu \vec \pi)^2
\right] - U(\sigma, \pi) \, , 
\end{eqnarray}
where $U(\sigma,\vec \pi)= \lambda^2 (\sigma^2+\vec
\pi^2-\nu^2)^2/8-f_\pi m_\pi^2 \sigma$ is the standard Mexican hat
potential implementing both spontaneous breaking of chiral symmetry as
well as PCAC yielding the Goldberger-Treiman relation $M_q = g_{\pi q
  q} f_\pi = g_{\sigma q q} f_\pi $ at the constituent quark
level. When this model is interpreted from a gradient expansion of the
NJL model quarks are regarded as valence quarks whereas kinetic meson
terms arise from the polarization of the Dirac sea and $m_\sigma^2 = 4
M_q^2 + m_\pi^2 $, which for $M_q=M_N/3=M_V/2$ yields
$m_\sigma=650-770 {\rm MeV}$. In the heavy constituent quarks limit
the model implies $1\pi$ and $1\sigma$ exchange potentials,
\begin{eqnarray}
V_{qq'}^{1\pi} (\vec r) &=& -\frac{g_{\pi qq}^2}{4 M_q^2} \tau_q \cdot
\tau_q' \int \frac{d^3 p}{(2\pi)^3} e^{i \vec p \cdot \vec r }
\frac{(\sigma_q \cdot p)( \sigma_{q'} \cdot p )}{p^2+m_\pi^2} \, , 
\nonumber \\ 
V_{qq'}^{1\sigma} (\vec r) &=&
  g_{\pi qq}^2 \int \frac{d^3 p}{(2\pi)^3} e^{i \vec p \cdot \vec r }
  \frac{1}{p^2+m_\sigma^2} =- \frac{g_{\pi qq}^2}{4 \pi}
  \frac{e^{-m_\sigma r}}{r} \, , 
\label{eq:vqq}
\end{eqnarray}
whence baryon properties can be obtained by solving the Hamiltonian
\begin{eqnarray}
H= \sum_{i=1}^{N_c} \left[ \frac{p_i^2}{2M_q} + M_q \right] +
\sum_{i < j} V(x_i - x_j) = \frac{P^2}{2M}+ N_c M_q + H_{\rm int} \, , 
\end{eqnarray}
where the total momentum $P = \sum_{i=1}^{N_c} p_i /N_c$ and the
intrinsic Hamiltonian have been introduced. Due to Galilean invariance
the wave function of a moving baryon can be factorized 
\begin{eqnarray}
\Psi_B (x_1 , \dots , x_{N_c} ) = \phi ( \xi_1 , \dots , \xi_{N_c-1} )
e^{iP \cdot R} \, , 
\end{eqnarray} 
with $R = \sum_{i=1}^{N_c} x_i /N_c$ the CM of the cluster and $\xi_i
= x_i -R/N_c$ intrinsic coordinates, $\sum_i \xi_i=0$. We will assume
that this complicated problem has been solved
already~Ref.~\cite{Glozman:1995fu}.  For large $N_c$ the Hartree mean
field approximation $\Psi_B (x_1 , \dots , x_{N_c} ) =
\prod_{i=1}^{N_c} \phi_{\alpha_i} ( x_i) \chi_c $ might be
used~\cite{Goity:2004pw}).  For separated hadrons the 
interaction between quark clusters A and B can be written as sum of
pairwise interactions which, for elementary $\pi qq$ and $\sigma
qq$ vertices, reads
\begin{eqnarray}
V_{\rm int} (\vec x_1 , \dots , \vec x_{N_c}; \vec y_1 , \dots , \vec y_{N_c}) =
\sum_{i,j} V_{ij}^{\sigma+\pi} (\vec x_i - \vec y_j) = \int \frac{d^3 q}{(2\pi)^3}
\sum_{i,j} V_{ij}^{\sigma + \pi} (q) e^{i \vec q \cdot (\vec x_i - \vec y_j)} \, .
\end{eqnarray}
Switching to intrinsic coordinates variables  $ \vec x_i
= \vec \xi_i + \vec R/2 $ and $ \vec y_j = \vec \eta_j - \vec R/2 $ with $
\sum_i \xi_i = \sum_j \eta_j = 0 $ where $R$ is the distance between
the CM of each cluster, we have
\begin{eqnarray}
V_{1\pi} (\vec R) &=& \frac{g_{\pi qq}^2}{M_q^2} \int \frac{d^3 q}{(2\pi)^3} 
e^{i q \cdot R} \, \frac{q_k q_k}{q^2+m_\pi^2} 
 G^{ka}_A (q) G^{ka}_B (q)^* \, , \\   
V_{1\sigma} (\vec R) &=& g_{\pi qq}^2  \int \frac{d^3 q}{(2\pi)^3} 
e^{i q \cdot R} \, \frac{1}{q^2+m_\sigma^2} 
 \rho_A (q) \rho_B (q)^* \, , 
\end{eqnarray}
where the spin-isospin density and scalar densities are given
by (e.g. cluster A)
\begin{eqnarray}
G^{ka}_A (q) &=& \frac12 \sum_{i=1}^{N_c} \sigma^k_i \tau^a_i e^{i \xi_i
  \cdot q} \, ,  \qquad \rho_A (q) = \frac1{N_c}\sum_{i=1}^{N_c} e^{i \xi_i \cdot q} \, , 
\end{eqnarray}
respectively. Note that the scalar and Baryon densities as well as the
pseudoscalar and axial densities coincide unlike the relativistic
case. That means that {\it within } the approximations one should have
$M_S=M_V$. Thus, the total Hamiltonian is written as
\begin{eqnarray}
H= H_{\rm A,int} + H_{\rm B,int} + V_{\rm int} (R) + \frac{P^2}{2 M_T}+ \frac{p^2}{2 \mu} \, . 
\end{eqnarray}
Galilean invariance implies that {\it inertial masses} are $M_T = 2
N_c M_q $ and $\mu = N_c M_q /2$. Introducing the two independent
cluster complete states $ H_{\rm A,int} \phi_{A,n} = M_{A,n}
\phi_{A,n} $ and $ H_{\rm B,int} \phi_{B,m} = M_{B,m} \phi_{B,m} $ the
two-clusters CM frame unperturbed wave function is just a product
\begin{eqnarray} 
\Psi^{(0)}_{A_n,B_m} (1,2,3;4,5,6) = \phi_{A,n} (1,2,3; R/2) \phi_{B,m} (4,5,6; -R/2) e^{iQ \cdot R} \, , 
\end{eqnarray}
where $Q $ is the relative momentum between the two clusters.  The
above problem is usually handled by Resonating Group
Methods~\cite{Oka:1984yw,AlvarezEstrada:1986wq,Valcarce:2005em,Bartz:2000vm}. We
analyze this coupled channel scattering problem perturbatively where
the transition potentials, defined as $ V_{A_n B_m;A_k B_l} (R) =
\langle \phi_{A,n} \phi_{B,m} | V_{\rm int} | \phi_{A,k} \phi_{B,l}
\rangle $, have a familiar folding structure which in the case of the
pion reads
\begin{eqnarray}
V_{A_n B_m;A_k B_l}^{1\pi} (R) = \frac{g_{\pi qq}^2}{M_q^2}\int
\frac{d^3 q}{(2\pi)^3} \frac{q_i q_j}{q^2+m_\pi^2} e^{i q \cdot R}
\langle A_n | G_{ia} (q) |A_k
\rangle \langle B_m | G_{ja} (-q) |B_l \rangle \, . 
\end{eqnarray}

\section{Long distance limit and the need for renormalization}

At long distances the leading singularities $q = i m_\pi$ and $q = i
m_\sigma$ dominate~\cite{Entem:2007jg,FernandezCarames:2008en}. Using
that $ |\langle N | \rho (q) | N \rangle |^2 $ is an even function of
$q$ we get the structure for the $NN \to NN$ potentials
\begin{eqnarray}
V_{\sigma} (\vec R) &=& g_{\pi qq}^2 N_c^2 \int \frac{d^3 q}{(2\pi)^3}
e^{i q \cdot R}   \frac{|\langle N | \rho ( i m_\sigma) | N
\rangle |^2}{q^2+m_\sigma^2}  + C_0 \, \delta^{(3)}(R) + C_2 (-\nabla^2+m_\sigma^2)
\delta^{(3)}(R) + \dots \nonumber \\ 
&=& - \frac{g_{\sigma NN}^2}{4 \pi}
\frac{e^{-m_\sigma r}}{r} + {\rm distributions} 
 \end{eqnarray}
and Eq.~(\ref{eq:pot-OPE}) for the OPE contribution.  Here, the
couplings are given by $ g_{\sigma NN} = N_c g_{\sigma qq} |\rho(i
m_\sigma)| $ and $ g_{\pi NN} = N_c g_A g_{\pi qq} |\rho(i m_\pi)| $
where $g_A = (N_c+2)/3$~\cite{Karl:1984cz}. Assuming $|\rho(i m_\pi)|
\sim |\rho(0)|=1$ one has the Goldberger-Treiman relation $g_A M_N =
g_{\pi NN} f_\pi$ at the nucleon level. Thus, at long distances finite
size effects are represented as an infinite sum of delta functions and
derivatives thereof. However, any finite truncation will produce a
negligible contribution at any non-vanishing distance. In a sense,
this result is reminiscent of the Gauss theorem for charged objects
with a sharp non-overlapping boundary; the interaction is mainly due
to the total charge and regardless on the density profiles of the
system. Only an infinite number of terms may yield a finite size
effect. Note that the coefficients of the contact interactions are
{\it fixed} numbers having a meaning perturbatively. However, if one
tries to play with them to characterize finite resolution effects
(nucleon size and potential range) in a model independent way
non-perturbatively (solving e.g. the Schr\"odinger equation) important
restrictions arise.  Unlike the $\delta's$, the OPE short distance
$1/r^3$ singularity is not located in a compact region, i.e. is not
killed by taking a finite support test function, and contributes to
all arbitrarily small distances. Thus, one can effectively drop the
derivatives of distributions.  This simple-minded argument was
advanced in Ref.~\cite{PavonValderrama:2005wv} and explicitly verified
in momentum space by taking $C_0$ and $C_2$ as {\it real} counterterms
in the Lippmann-Schwinger equation in Ref.~\cite{Entem:2007jg}; either
$C_2$ becomes irrelevant or the scattering amplitude does not
converge. Therefore, we represent $C_0$ as an energy independent
boundary condition. The renormalization procedure in coordinate space
generally corresponds to 1) fix some low energy constants such as
e.g. the scattering length for s-waves, $\alpha_0$, at zero energy as
an {\it independent} variable of the potential, 2) integrate in down
to an arbitrarily small cut-off radius $r_c$, 3) construct an
orthogonal finite energy state by matching log-derivatives at $r_c$
and 4) integrating out generating a phase-shift $\delta_0(p)$ with a
{\it prescribed} scattering length $\alpha_0$. This prescription is
the {\it renormalization condition} and the procedure of integrating
in and out corresponds to evolving along the renormalization
trajectory.  The crucial aspect is that short distance insensitivity
is implemented.  The $\pi+\sigma$ model and OBE extensions are
analyzed in detail in
Refs.~\cite{RuizArriola:2007wm,CalleCordon:2008eu,Cordon:2009pj} where
form factors {\it after renormalization} are found to be marginal.

\section{Renormalization of Spin-flavour Van der Waals forces}

\begin{figure*}[t]
\begin{center}
\epsfig{figure=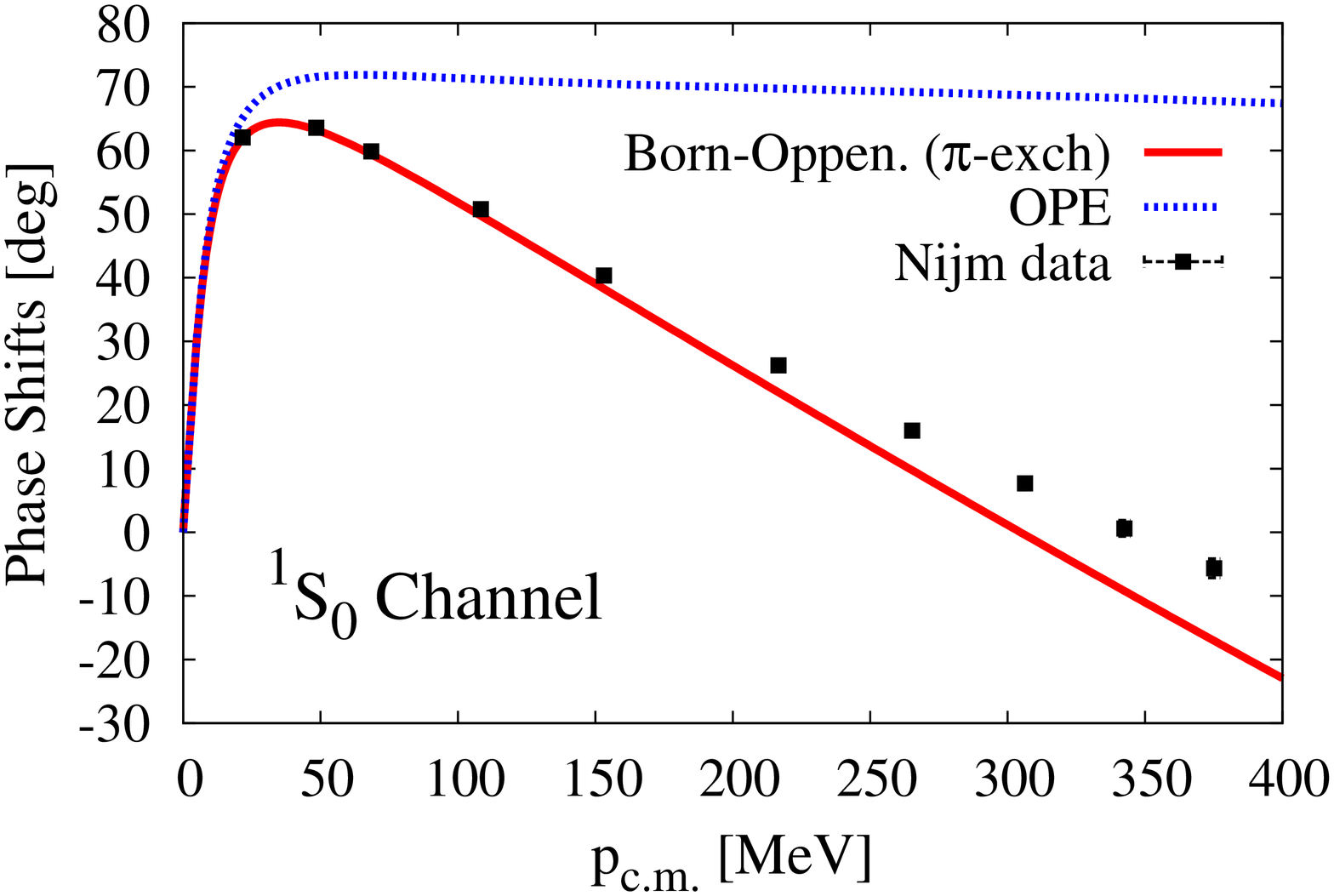,height=6.9cm,width=5cm,angle=270}
\epsfig{figure=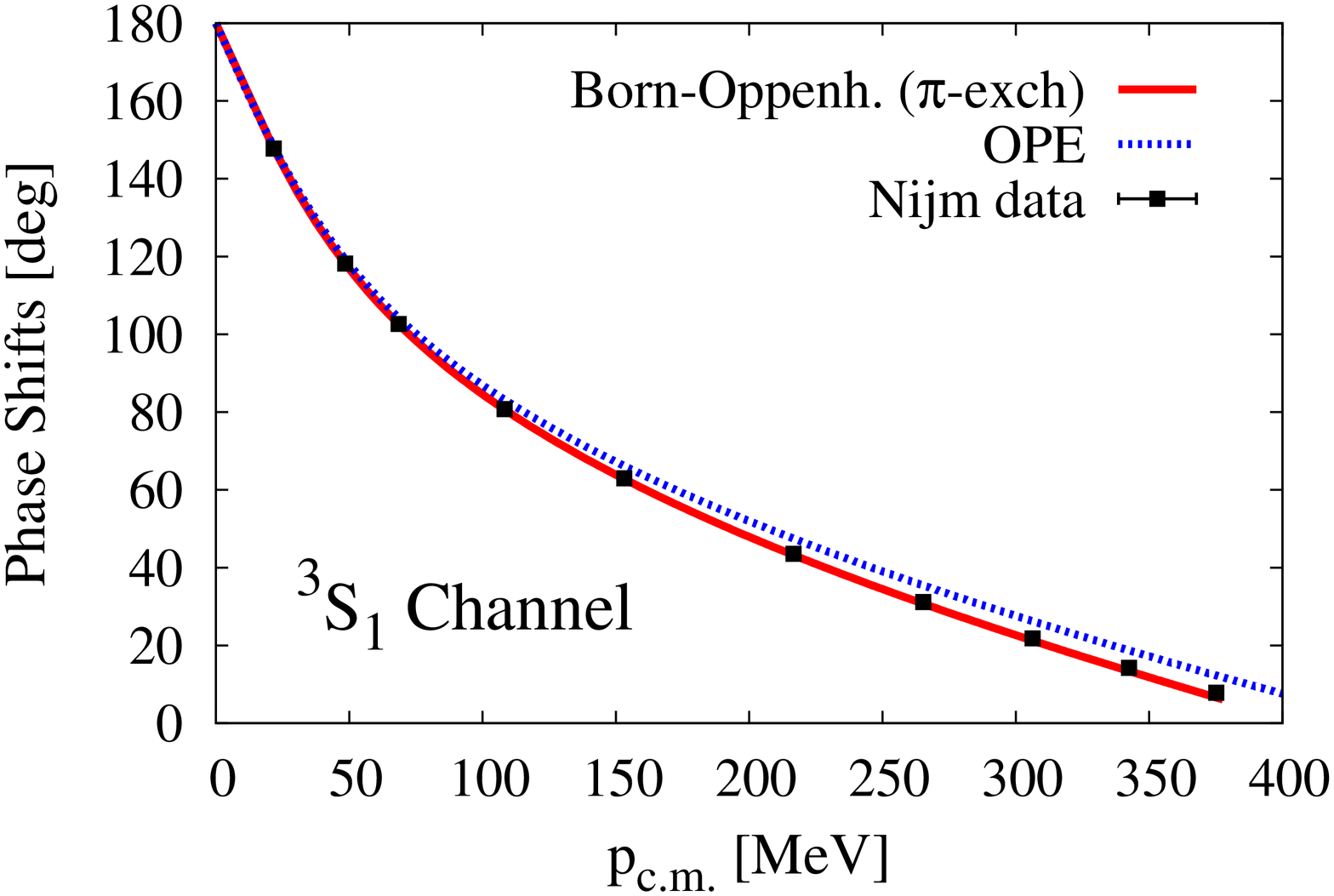,height=6.9cm,width=5cm,angle=270} \\ 
\epsfig{figure=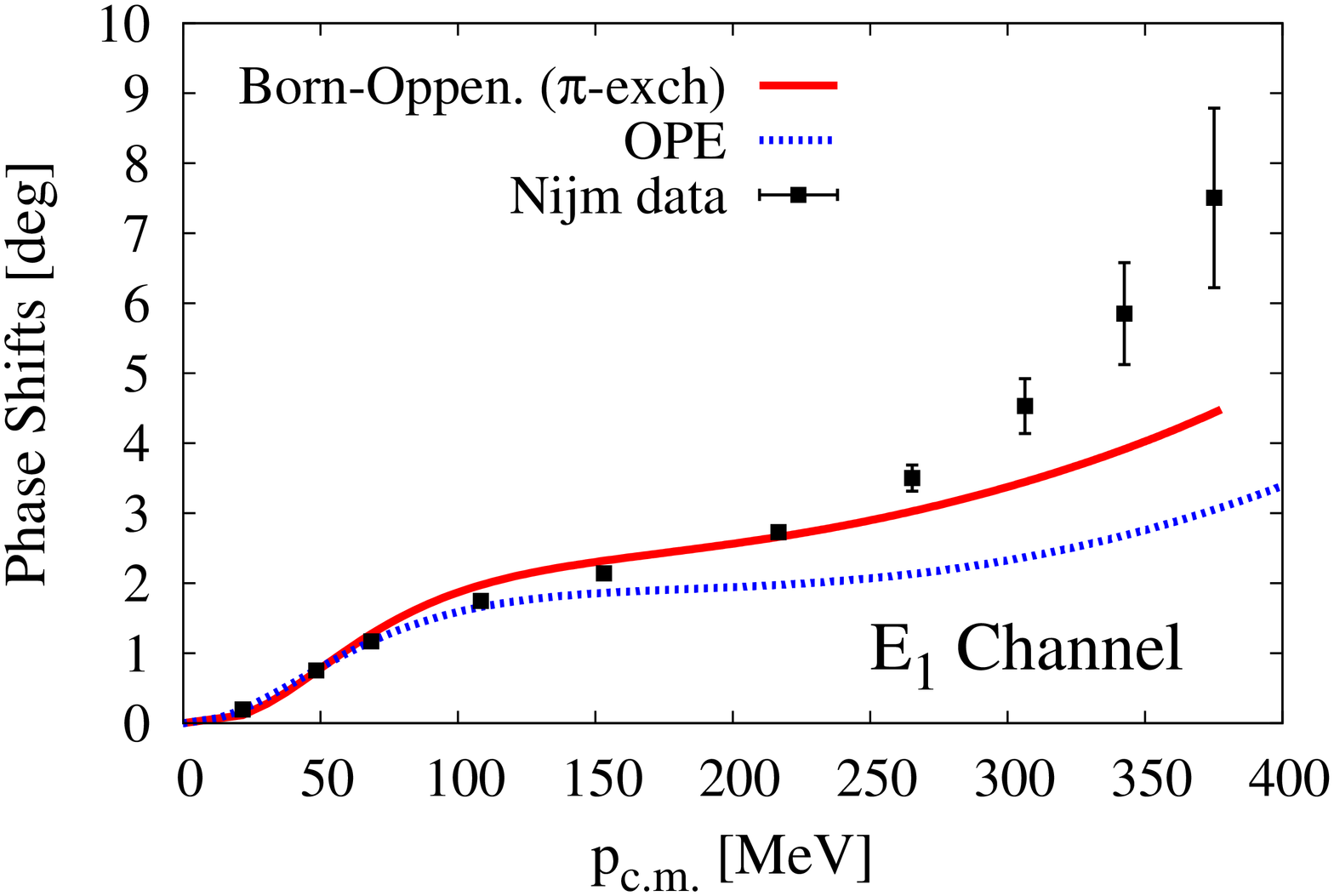,height=6.9cm,width=5cm,angle=270}
\epsfig{figure=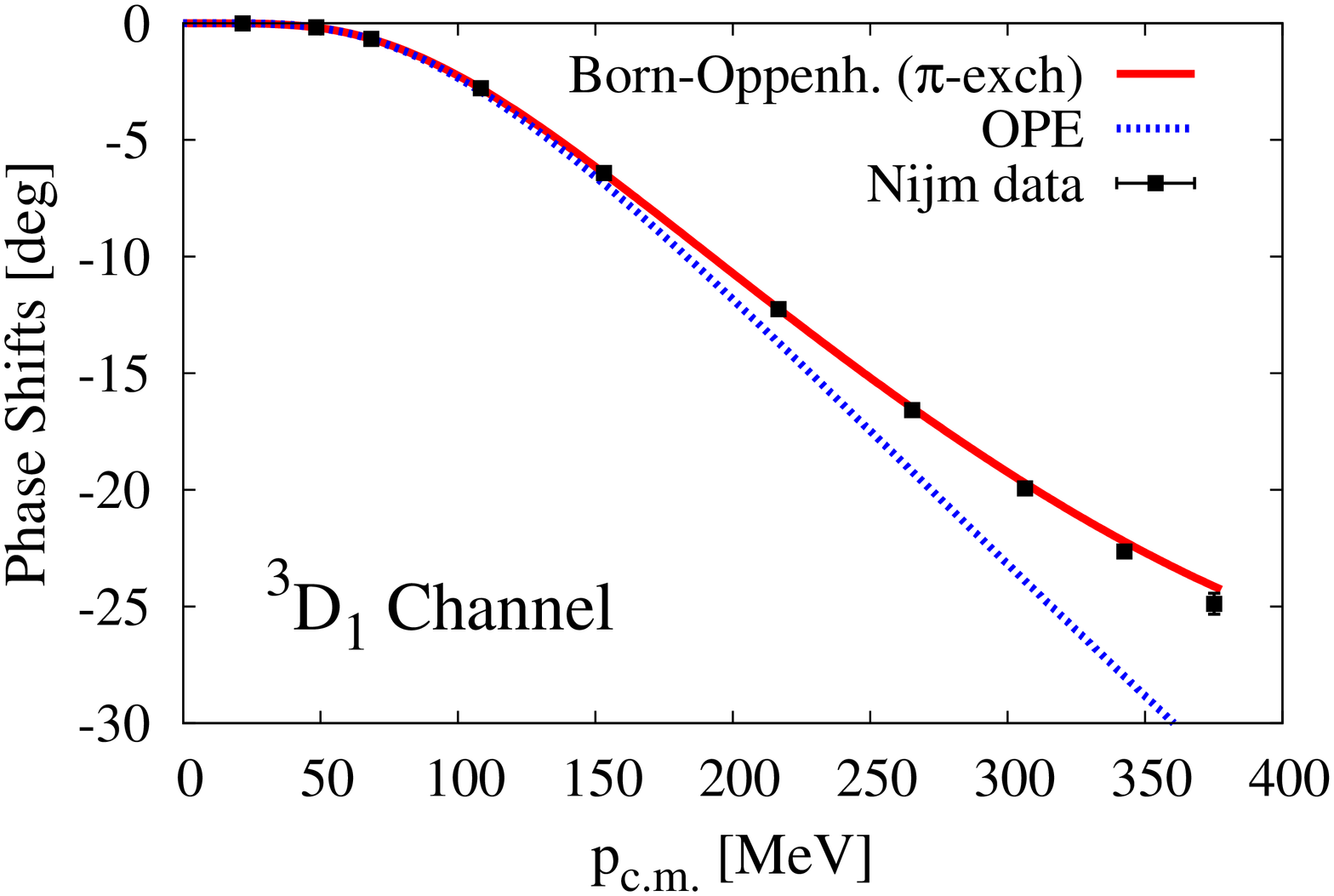,height=6.9cm,width=5cm,angle=270}
\end{center}
\caption{\footnotesize Renormalized (eigen) phase shifts for the OPE and
  $\Delta$-Born-Oppenheimer potentials as a function of the CM np
  momentum $p$ in the spin singlet $^1S_0$ (one counterterm) and
  triplet $^3S_1-^3D_1$ (three counterterms) channels compared to
  averaged Nijmegen potentials~\cite{Stoks:1994wp}. We take $f_{\pi 
    NN}^2/4\pi=0.07388$\cite{Stoks:1994wp} and $f_{\pi N
    \Delta}/f_{\pi NN} = 6\sqrt{2}/5$.}
\label{fig:phase-shift_BO}
\end{figure*}

The non-linear chiral quark model~\cite{Manohar:1983md} corresponds to take 
$m_\sigma \to \infty$, reducing to just OPE.  The results for
the phase shifts in the lowest partial waves are presented in
Fig.~\ref{fig:phase-shift_BO}.  Note the bad $^1S_0$ phase. To 
improve on this the long distance OPE transition potential is taken 
\begin{eqnarray}
V_{AB;CD} (R)= (\vec \tau_{AB} \cdot \vec \tau_{CD}) \Big\{
\sigma_{AB} \cdot \sigma_{CD} [W_S^{1\pi}]_{AB;CD} (R)+
      [S_{12}]_{AB;CD} [W_T^{1\pi}]_{AB;CD} (R)\Big\} \, , 
\end{eqnarray}
where the tensor term is defined as $S_{12}= 3 (\sigma_{AB} \cdot \hat R
)(\sigma_{CD} \cdot \hat R )- \sigma_{AB} \cdot \sigma_{CD} $ and
\begin{eqnarray}
[W_{S,T}^{1\pi} ]_{AB;CD} (R) = \frac{m_\pi}{3} \frac{f_{\pi AC} f_{\pi
    BD}}{4\pi} Y_{0,2}(m_\pi R) 
\end{eqnarray}
Note that also here there is a $1/r^3$ singularity. In this particular
form the resulting potential is model
independent~\cite{Green:1976wx}~\footnote{The corresponding couplings
  are $f_{\pi AB} = |F_{\pi AB} (i m_\pi)|$ where the transition form
  factors are defined as $ F_{\pi AB} (q^2) \chi^\dagger_A T^a S^i
  \chi_B = \langle A | G^{ia} (q) |B \rangle $. In the $SU(4) \otimes
  SU_c(N_c) $ quark model~\cite{Karl:1984cz} and in the chiral limit
  they fulfill $f_{\pi \Delta \Delta}/f_{\pi NN}= 1/5 $ and $f_{\pi N
    \Delta}/f_{\pi NN}= 3[(N_c-1)(N_c+5)/2]^\frac12/ (N_c+2)$. The
  $\Delta \to N \pi$ width in the Born approximation yields $f_{\pi N
    \Delta}^2/(4\pi)= 0.324$.}. 
In general, this requires solving a coupled channel
problem~\cite{Niephaus:1979mw,Wiringa:1984tg} but if we are
interested in the elastic channel with $T_{\rm CM}= m_\pi < \Delta
\equiv M_\Delta-M_N = 293 {\rm MeV}$ we may take into account the
effect of the closed channels as sub-threshold effects in perturbation
theory. We neglect the exponentially $\sim e^{-2 M_q r}$ suppressed
quark exchange contribution. In obvious operator-matrix notation and
restricting to the two particle ground $|0\rangle = | NN\rangle$ and
excited $|n\rangle = | N \Delta\rangle,| \Delta N \rangle,|
\Delta\Delta\rangle$ in-going and out-going channels and resolvent
$G_{0,k}(E)=(E-H_{0,k})^{-1}$ with $H_{0,k}=P^2/(2\mu_k)+E_k$, we get
for the T-matrix
\begin{eqnarray}
(T)_{nm} = (V)_{nm} + \sum_k (V)_{nk} G_{0,k} (V)_{k,m} + {\cal O} (V^3) \, , 
\end{eqnarray}
with $E_0=2 M_N$,$E_{1,2}=M_N+M_\Delta$ and $E_{3}= 2M_\Delta$ the
corresponding thresholds. Thus, separating the elastic term $k=0$
explicitly from the sum we get the effective potential in the elastic
scattering channel corresponding to higher pion exchanges,
wich, when iterated to second order yields the elastic scattering
amplitude $T_{00}$.  Specifically, defining the momentum space
potential $V_{nm} (k'-k) \equiv \langle k' , n | V | k , m \rangle =
\int d^3 R V_{nm} (R) e^{i(k-k') \cdot R}$ we get
\begin{eqnarray}
\bar V_{00} (k'-k,E) =V_{00} (k'-k) + \sum_{n\neq 0} \int \frac{d^3
  q}{(2\pi)^3}  \frac{V_{0n} (k'-q)V_{n0} (q-k)}{E-q^2/2\mu_n - E_n } + {\cal O}(V^3)
\end{eqnarray}
which, expectedly, depends on the energy. Evaluating on-shell at $E=
E_0 + p^2/2\mu_0$, assuming a large splitting $p \ll \sqrt{\Delta
  M_\Delta} = 600 {\rm MeV} $ and neglecting the kinetic energy piece
in the $N \Delta$ cannel, we get the perturbative and local optical
potential in coordinate space
\begin{eqnarray}
\bar V_{NN;NN}^{1\pi + 2 \pi + \dots } (R) &=&  V^{1\pi}_{NN,NN} (R) + 
\frac{ 2 |V^{1\pi}_{NN,N\Delta} (R) 
  |^2}{ M_{N}-M_{\Delta}}+ {\cal O}(V^3) 
\label{eq:oppenheimer} 
\end{eqnarray} 
which is the Born-Oppenheimer approximation to second order which
generates more complicated spin-isospin structures than just OPE {\it
  including} a central force, all of them $\sim e^{-2 m_\pi R}$
and resembling TPE. Note that only the intermediate $N\Delta$ state
contributes. The above result implies an attractive and short distance
singular potential since $V^{1\pi}_{NN,N\Delta} (R) \sim g_A^2 /(
f_\pi^2 R^3)$ and hence the potential becomes singular $ \bar
V_{NN,NN} \sim -g_A^4 / ( \Delta f_\pi^4 R^6) $.
Actually, Eq.~(\ref{eq:oppenheimer}) was evaluated in the Skyrme
soliton model within the Heitler-London approximation, i.e. the
product ansatz in the coupled channel
space~\cite{Walet:1992zza,Walet:1992gw} providing the long sought mid
range attraction~\cite{Walhout:1992ek}.~\footnote{Molecular methods
  used in the Skyrme
  model~\cite{Walet:1992zza,Walet:1992gw,Walhout:1992ek} are replaced
  by evaluating model form factor yielding regularized Meson Exchange
  potentials~\cite{Holzwarth:1996bc} where the only remnant of the
  model is in the meson-form factors.} We reproduce the same results
in the quark model calculation. The potential found using Feynman
graph techniques~\cite{Kaiser:1998wa} looks very similar with
identical short distance singular behaviour identifying $h_A /g_A=
f_{\pi N \Delta} / 2 f_{\pi N N}$. Note that we leave out background
$\pi N$ scattering which correspond to triangle and box TPE diagrams
at the quark level.  The renormalization procedure as well as the
necessary counterterms in the general coupled channel singular
potentials has been explained in much detail in
Ref.~\cite{PavonValderrama:2005wv,PavonValderrama:2007nu}. The results
for the phase shifts using Eq.~(\ref{eq:oppenheimer}) in the lowest
partial waves are depicted in Fig.~\ref{fig:phase-shift_BO}. In any
case the description looks extremely similar (including deuteron
properties) to the renormalization~\cite{Valderrama:2008kj} of more
sophisticated field theoretical
potentials~\cite{Kaiser:1998wa}. Convergence is achieved already at
$r_c \sim 0.5 {\rm fm}$.

The multiplicative structures of Eq.~(\ref{eq:oppenheimer}) reflect
spin-flavour excitations and remind of the analogous Van der Waals
forces in atomic systems. They hold literally even after inclusion of
form factors with folded potentials (although $\Lambda_{\pi NN}$,
$\Lambda_{\pi N\Delta}$ and $\Lambda_{\pi \Delta\Delta}$ are not
necessarily identical) which remove the singularity. This is {\it not}
equivalent to regularize the effective potential as a whole through
subtractions.  We have checked that form factors {\it
  after renormalization} become marginal in agreement with the OBE
analysis~\cite{Cordon:2009pj}.

\section{Wigner SU(4) as a long distance symmetry}

If the tensor force component of the qq potential, Eq.~(\ref{eq:vqq}),
is neglected one has invariance under the spin-isospin $SU(4)$ group
with the quarks in the fundamental ${\bf 4}$-dimensional
representation, $q=(u \uparrow, u \downarrow, d \uparrow, d
\downarrow)$.  In the three quark system we have the spin-flavour
states $ {\bf 4} \otimes {\bf 4} \otimes {\bf 4} = {\bf 4}_A \oplus
{\bf 20}_S \oplus {\bf 20}_{M_1} \oplus {\bf 20}_{M_2} $.  Due to
colour antisymmetry only the symmetric state survives which
spin-isospin, $(S,T)$, decomposition is ${\bf 20}_S= (\frac12,\frac12)
\oplus (\frac32,\frac32)= N \oplus \Delta $ yielding $N-\Delta$
degeneracy. Since $M_\Delta - M_N $ is large at nuclear scales, one
might still treat the Nucleon quartet $N=(p \uparrow, p \downarrow, n
\uparrow, n \downarrow)$ as the fundamental rep.  of the old
Wigner-Hund $SU(4)$ symmetry which implies spin independence, in
particular that $V_{^1S_0}(r)=V_{^3S_1}(r)$ at {\it all distances}
suggesting that phases $\delta_{^1S_0}(p)=\delta_{^3S_1}(p)$ in
contradiction to data (see e.g. Fig.~\ref{fig:phase-shift_BO}). The
amazing finding of Ref.~\cite{CalleCordon:2008cz} was that assuming
identical potentials $V_{^1S_0}(r)=V_{^3S_1}(r)$ for $r > r_c \to 0$
one has
\begin{eqnarray}
p \cot \delta_{^1S_0} (p) = \frac{ \alpha_{^1S_0} {\cal A} ( p) +
{\cal B} (p)}{ \alpha_{^1S_0} {\cal C} ( p) + {\cal D} (p)} \, , \qquad p
\cot \delta_{^3S_1} (p) = \frac{ \alpha_{^3S_1} {\cal A} ( p) + {\cal
B} (p)}{ \alpha_{^3S_1} {\cal C} ( p) + {\cal D} (p)} \, , 
\end{eqnarray} 
where the functions ${\cal A}(p)$, ${\cal B}(p)$, ${\cal C}(p)$ and
${\cal D}(p)$ are {\it identical} in both channels, but the
experimentally different scattering lengths $\alpha_{^1S_0}= -23.74
{\rm fm} $ and $ \alpha_{^3S_1}= 5.42 {\rm fm} $ yield quite different
phase shifts with a fairly good agreement. Thus, Wigner symmetry is
broken by very short distance effects and hence corresponds to a {\it
  long distance symmetry} (a symmetry broken only by counterterms).
Moreover, large $N_c$~\cite{Kaplan:1996rk} suggests that Wigner
symmetry holds only for {\it even} L, a fact verified by phase shift
sum rules~\cite{CalleCordon:2008cz}. In
Refs.~\cite{CalleCordon:2009ps,RuizArriola:2009bg} we analyze further
the relation to the old Serber symmetry which follows from vanishing
P-waves in $S=1$ channels, showing how old nuclear symmetries are
unveiled by coarse graining the NN interaction via the $V_{\rm low k}$
framework~\cite{Bogner:2003wn} and with testable implications for
Skyrme forces in mean field calculations~\cite{Zalewski:2008is}.

The chiral quark model is supposedly an approximate non-perturbative
description, but {\it perturbative} gluons may be introduced by
standard minimal coupling~\cite{Valcarce:2005em}, $i
\slashchar{\partial} \to i \slashchar{\partial}+g \slashchar{A}^a
\cdot {\lambda^c_a}/2 $ with $\lambda^c_a$ the $N_c^2-1$ Gell-Mann
      {\it colour } matrices. A source of $SU(4)$ breaking is the
      contact one gluon exchange which yields spin-colour
      chromo-magnetic interactions ($S_{ij}$ is the tensor operator),
\begin{eqnarray}
V^{\rm OGE} = \frac14 \alpha_s  \sum_{i<j} (\lambda_i^c \cdot \lambda_j^c)
\left\{ \frac1{r_{ij}} 
- \frac{\pi}{4m_im_j}\left[1+ \frac23 \sigma_i \cdot \sigma_j \right] 
\delta^{(3)}(\vec r_{ij})  - \frac{3}{4m_i m_j r_{ij}} S_{ij} \right\} 
\end{eqnarray}
breaking the $\Delta-N$ degeneracy. This short distance terms break
{\it also} the $^1S_0$ and $^3S_1$ degeneracy of the $NN$ system
providing an understanding of the long distance character of Wigner
symmetry.  Taking the Wigner symmetric zero energy state and
perturbing around it, the previous argument suggests that
$1/\alpha_{^3S_1}- 1/\alpha_{^1S_0} = {\cal O}(M_\Delta-M_N)$ with a
computable coefficient.

\section{Conclusions}

Chiral Quark and Soliton models while quite different in appearance provide
some universal behaviour regarding $NN$ interactions. If the
asymptotic potentials coincide, the main differences in describing the
scattering data are due to a few low energy constants which in some cases
are subjected to extreme fine tuning of the model parameters. The
success of the model at finite energy is mainly reduced to reproducing
these low energy parameters.

\vskip1cm 

{\em One of us (E.R.A.)  warmly thanks M. Rosina, B. Golli and
  S.~\v{S}irca for the invitation and D. R. Entem, F. Fernandez,
  M. Pav\'on Valderrama and J. L. Goity for discussions.  This work is
  supported by the Spanish DGI and FEDER funds with grant
  FIS2008-01143/FIS, Junta de Andaluc{\'\i}a grant FQM225-05, and EU
  Integrated Infrastructure Initiative Hadron Physics Project contract
  RII3-CT-2004-506078.}


\newpage 


\begin{thebibliography}{10}

\bibitem{Machleidt:1987hj}
R. Machleidt, K. Holinde and C. Elster,
\newblock Phys. Rept. 149 (1987) 1.

\bibitem{Nagels:1977ze}
M.M. Nagels, T.A. Rijken and J.J. de~Swart,
\newblock Phys. Rev. D17 (1978) 768.

\bibitem{PavonValderrama:2005gu}
M. Pavon~Valderrama and E. Ruiz~Arriola,
\newblock Phys. Rev. C72 (2005) 054002.

\bibitem{RuizArriola:2007wm}
E. Ruiz~Arriola, A. Calle~Cordon and M. Pavon~Valderrama,
\newblock (2007), 0710.2770.

\bibitem{CalleCordon:2008eu}
A. Calle~Cordon and E. Ruiz~Arriola,
\newblock AIP Conf. Proc. 1030 (2008) 334. 

\bibitem{CalleCordon:2008cz}
A. Calle~Cordon and E. Ruiz~Arriola,
\newblock Phys. Rev. C78 (2008) 054002.

\bibitem{CalleCordon:2009ps}
A. Calle~Cordon and E. Ruiz~Arriola,
\newblock Phys. Rev. C80 (2009) 014002.

\bibitem{RuizArriola:2009bg}
E. Ruiz~Arriola and A. Calle~Cordon,
\newblock (2009), 0904.4132.

\bibitem{Cordon:2009pj}
A. Calle~Cordon and E. Ruiz~Arriola,
\newblock (2009), 0905.4933.

\bibitem{Oka:1984yw}
M. Oka and K. Yazaki,
\newblock Int. Rev. Nucl. Phys. 1 (1984) 489.

\bibitem{AlvarezEstrada:1986wq}
R.F. Alvarez-Estrada, F. Fernandez, J. L. Sanchez-Gomez and V. Vento,
\newblock Lect. Notes Phys. 259 (1986) 1.

\bibitem{Walhout:1992ek}
T.S. Walhout and J. Wambach,
\newblock Int. J. Mod. Phys. E1 (1992) 665.

\bibitem{Valcarce:2005em}
A. Valcarce, H. Garzilazo, F. Fernandez and P. Gonzalez,
\newblock Rept. Prog. Phys. 68 (2005) 965.

\bibitem{Ishii:2006ec}
N. Ishii, S. Aoki and T. Hatsuda,
\newblock Phys. Rev. Lett. 99 (2007) 022001.

\bibitem{Aoki:2009ji}
S. Aoki, T. Hatsuda and N. Ishii,
\newblock (2009), 0909.5585.

\bibitem{O'Connell:1995wf}
H.B. O'Connell et~al.,
\newblock Prog. Part. Nucl. Phys. 39 (1997) 201.

\bibitem{Stoks:1994wp}
V.G.J. Stoks et~al.,
\newblock Phys. Rev. C49 (1994) 2950.

\bibitem{Gavela:1979zu}
M.B. Gavela et~al.,
\newblock Phys. Lett. B82 (1979) 431.

\bibitem{Greenberg:1981xn}
O.W. Greenberg and H.J. Lipkin,
\newblock Nucl. Phys. A370 (1981) 349.

\bibitem{Christov:1995vm}
C.V. Christov et~al.,
\newblock Prog. Part. Nucl. Phys. 37 (1996) 91.

\bibitem{Weigel:2008zz}
H. Weigel,
\newblock Lect. Notes Phys. 743 (2008) 1.

\bibitem{Arriola:2006ds}
E. Ruiz~Arriola, W. Broniowski and B. Golli,
\newblock Phys. Rev. D76 (2007) 014008.

\bibitem{Kaplan:1996rk}
D.B. Kaplan and A.V. Manohar,
\newblock Phys. Rev. C56 (1997) 76.

\bibitem{Birse:1983gm}
M.C. Birse and M.K. Banerjee,
\newblock Phys. Lett. B136 (1984) 284.

\bibitem{Manohar:1983md}
A. Manohar and H. Georgi,
\newblock Nucl. Phys. B234 (1984) 189.

\bibitem{Glozman:1995fu}
L.Y. Glozman and D.O. Riska,
\newblock Phys. Rept. 268 (1996) 263.

\bibitem{Goity:2004pw}
J.L. Goity,
\newblock Phys. Atom. Nucl. 68 (2005) 624.

\bibitem{Bartz:2000vm}
D. Bartz and F. Stancu,
\newblock Phys. Rev. C63 (2001) 034001.

\bibitem{Entem:2007jg}
D. R. Entem, E. Ruiz~Arriola, M. Pavon~Valderrama and R. Machleidt,
\newblock Phys. Rev. C77 (2008) 044006. 

\bibitem{FernandezCarames:2008en}
M.T. Fernandez-Carames, P. Gonzalez and A. Valcarce,
\newblock Phys. Rev. C77 (2008) 054003.

\bibitem{Karl:1984cz}
G. Karl and J.E. Paton,
\newblock Phys. Rev. D30 (1984) 238.

\bibitem{PavonValderrama:2005wv}
M. Pavon~Valderrama and E. Ruiz~Arriola,
\newblock Phys. Rev. C74 (2006) 054001.

\bibitem{Green:1976wx}
A.M. Green,
\newblock Rept. Prog. Phys. 39 (1976) 1109.

\bibitem{Niephaus:1979mw}
G.H. Niephaus, M. Gari and B. Sommer,
\newblock Phys. Rev. C20 (1979) 1096.

\bibitem{Wiringa:1984tg}
R.B. Wiringa, R.A. Smith and T.L. Ainsworth,
\newblock Phys. Rev. C29 (1984) 1207.

\bibitem{Walet:1992zza}
N.R. Walet, R.D. Amado and A. Hosaka,
\newblock Phys. Rev. Lett. 68 (1992) 3849.

\bibitem{Walet:1992gw}
N.R. Walet and R.D. Amado,
\newblock Phys. Rev. C47 (1993) 498.

\bibitem{Holzwarth:1996bc}
G. Holzwarth and R. Machleidt,
\newblock Phys. Rev. C55 (1997) 1088.

\bibitem{Kaiser:1998wa}
N. Kaiser, S. Gerstendorfer and W. Weise,
\newblock Nucl. Phys. A637 (1998) 395.

\bibitem{PavonValderrama:2007nu}
M. Pavon~Valderrama and E. Ruiz~Arriola,
\newblock Annals Phys. 323 (2008) 1037.

\bibitem{Valderrama:2008kj}
M. Pavon~Valderrama and E. Ruiz Arriola, 
\newblock Phys. Rev. C79 (2009) 044001.


\bibitem{Bogner:2003wn}
  S.~K.~Bogner, T.~T.~S.~Kuo and A.~Schwenk,
\newblock  Phys.\ Rept.\  386 (2003) 1

\bibitem{Zalewski:2008is}
  M.~Zalewski, J.~Dobaczewski, W.~Satula and T.~R.~Werner,
\newblock  Phys.\ Rev.\  C77 (2008) 024316



\end{thebibliography}
\end{document}